\documentclass[pss]{wiley2sp} 
\usepackage{amsmath}
\usepackage{amssymb}

\begin{document}

\title{Organic glasses: cluster structure of the random energy landscape and its effect on charge transport and injection}

\titlerunning{Organic glasses: cluster structure, charge transport and injection}

\author{%
  Sergey V. Novikov}

\authorrunning{Sergey V. Novikov}

\mail{e-mail
  \textsf{cnovikov@gmail.com}, Phone:
  +7-495-9522428, Fax: +7-495-9525308}

\institute{%
  A.N. Frumkin Institute of Physical Chemistry and
Electrochemistry, Leninsky prosp. 31, Moscow 119991, Russia}

\received{21 July 2010, revised 4 October 2010, accepted 4 October 2010} 
\published{11 March 2011} 


\abstract{%
%
%
%
\abstcol{%
  An appropriate model for the random energy landscape in organic
glasses is a spatially correlated Gaussian field, generated by randomly located and oriented dipoles and quadrupoles. Correlation properties of energetic disorder directly dictates the mobility dependence on the applied electric field. Electrostatic disorder is significantly modified in the vicinity of the electrode that affects injection }%
{ properties. Correlated Gaussian field forms clusters. We suggest a simple method to estimate an asymptotics of the cluster distribution on size for deep clusters where a value of the field on
each site is much greater than the rms disorder. Hopping transport in organic glasses in the case of high carrier density could be described in terms of the effective density-dependent temperature.}}

%
%

\maketitle   

\section{Introduction}
\label{sect_intro}
Most amorphous organic materials could be considered as organic glasses, and they belong to a very particular class of glasses. They are molecular glasses having very low concentration of intrinsic movable charge carriers. Hence, most organic glasses demonstrate measurable conductivity only after injection of carriers by the action of the laser pulse or strong electric field. At the same time, organic glasses usually have high concentration of molecular dipoles and quadrupoles. Due to almost zero concentration of free carriers and lack of screening, such molecules provide long range electrostatic contribution to the overall energetic disorder for charge carriers. Usually, the resulting random energy landscape $U(\vec{r})$ could be accurately approximated by the random Gaussian field with the typical magnitude of energetic disorder $\sigma=\left<U^2\right>^{1/2}$ close to $0.1$ eV \cite{Bassler:15,Borsenberger:9}. Long range sources inevitably lead to the strong spatial correlation of $U(\vec{r})$: in organic polar materials (dipolar glasses) correlation function $C(\vec{r}) =\left<U(\vec{r})U(0)\right>$ decays as $1/r$ and in nonpolar materials (quadrupolar glasses) it decays as $1/r^3$ \cite{Novikov:14573,Novikov:361}.
Long range correlations mean that sites with close values of random energy tend to group together and form clusters (cluster is defined as a set of connected sites,  where all of them have site energy $U$  greater than some boundary energy $U_0$) \cite{Novikov:14573}.
In this paper we consider how the correlated nature of organic glasses affects the important physical properties of these materials, namely charge carrier transport and injection.

\section{Charge transport in correlated Gaussian energy landscape}
\label{sect_transport}

Correlation nature of the energy landscape $U(\vec{r})$ directly dictates major features of the charge transport and injection in organic glasses. For example, the dependence of the quasi-equilibrium mobility $\mu$ on the applied electric field $E$ can be understood from the following simple consideration (for more thorough consideration see Ref. \cite{Dunlap:437}). Suppose that the carrier is located at the bottom of the potential well with the energy $U(0)$. Mobility is determined by the typical time for the carrier to reach a saddle point with the energy $U(\vec{r})-e\vec{E}\vec{r}$. That time can be estimated as
\begin{equation}
\label{typical_time}
t\simeq t_0 \exp\left[\frac{U(\vec{r})-U(0)-e\vec{E}\vec{r}}{kT}\right],
\end{equation}
where $1/t_0$ is an attempt frequency, and the average time for the Gaussian random field $U(\vec{r})$ is
\begin{eqnarray}
\label{typical_time_avg}
\left<t\right>\simeq t_0 \left<\exp\left[\frac{U(\vec{r})-U(0)}{kT}\right]\right>\exp\left(-\frac{e\vec{E}\vec{r}}{kT}\right)= \\ \nonumber
=t_0\exp\left\{\frac{\left<\left[U(\vec{r})-U(0)\right]^2\right>}{2(kT)^2}-\frac{e\vec{E}\vec{r}}{kT}\right\}= \\ \nonumber
=t_0 \exp\left[\frac{C(0)-C(\vec{r})}{(kT)^2}-\frac{e\vec{E}\vec{r}}{kT}\right].
\end{eqnarray}
Assuming a 1D transport in the dipolar glass (DG) with $C(\vec{r})=\sigma^2a/r$ ($a$ is a microscopic length, comparable to the typical distance between neighbor molecules \cite{Novikov:14573,Dunlap:542}), we can calculate the critical size of the potential well that provides the maximal escape time
\begin{equation}
\label{critical}
\frac{d\left<t\right>}{dr}=0, \hskip10pt r_{\rm cr}=\sigma
\left(\frac{a}{eEkT}\right)^{1/2},
\end{equation}
and the mobility is estimated as
\begin{equation}
\label{mob_crit}
\mu\propto 1/\left<t\left(r_{\rm cr}\right)\right>\propto \exp\left[-\left(\frac{\sigma}{kT}\right)^2+2\frac{\sigma}{kT}\left(\frac{eaE}{kT}\right)^{1/2}\right].
\end{equation}
This result provides a leading asymptotics of the exact solution of 1D transport problem \cite{Dunlap:542,Parris:5295}.
Characteristic mobility field dependence $\ln\mu\propto \sqrt{E}$ agrees well with the experimental data \cite{Borsenberger:9}. Extensive computer simulation of the 3D transport generally confirms Eq. (\ref{mob_crit}) and only modifies numeric parameters in this relation. Results of the simulation may be summarizes as a phenomenological relation
\begin{equation}
\ln\frac{\mu}{\mu _{0}}= -\left( \frac{3\sigma }{5kT} \right)
^{2} +C_{0} \left[ \left( \frac{\sigma }{kT} \right) ^{3/2}
-\Gamma \right] \sqrt{eaE/\sigma }  \label{eq5}
\end{equation}
where $C_0 \approx 0.78$, and $\Gamma \approx 2$ \cite{Novikov:4472}.

\begin{figure}[floatfix]
\begin{center}
\includegraphics[width=2.8in]{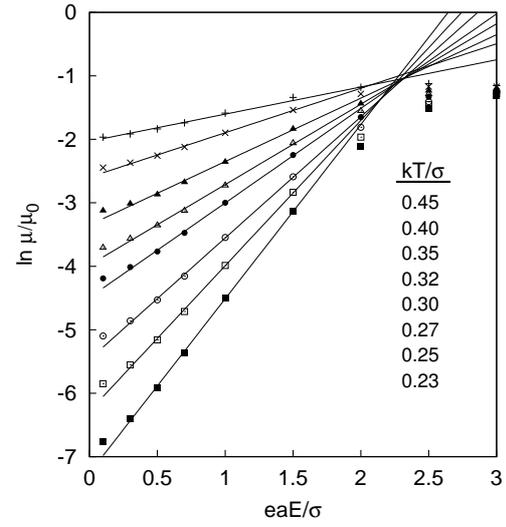}
\end{center}
\caption{Field dependent mobility in the GDM for different values
of $kT/\sigma$ (from the top curve downward). Straight lines show best fits for the linear regions of the curves.}
\label{GDM_fig}
\end{figure}

For the non-correlated Gaussian landscape (the Gaussian Disorder Model (GDM) \cite{Bassler:15}), mobility is controlled by the carrier release from the deep states to the neighbor sites having higher energy, hence the shift of the carrier energy in the applied field lead to the linear field dependence
\begin{equation}
\ln\mu\propto eaE/kT.  \label{mu_GDM_simple}
\end{equation}
This estimation is in good agreement with the simulation data (Fig. \ref{GDM_fig}); details of the simulation are similar to those described in Ref. \cite{Novikov:4472}. The overall mobility dependence on $T$ and $E$ could be with reasonable accuracy described as
\begin{equation}
\ln\frac{\mu}{\mu _{0}}= -0.38\left( \frac{\sigma }{kT} \right)
^{2} +1.17 \left( \frac{\sigma}{kT}-2.05
 \right) \frac{eaE}{\sigma}.  \label{mu_GDM}
\end{equation}

Mobility field dependence is directly governed by the spatial decay of the correlation function. For example, in non-polar organic materials the correlation function decays as $C\propto 1/r^3$ and $\ln\mu\propto E^{3/4}$ \cite{Novikov:361,Novikov:954}. In principle, the time-of-flight experiment (direct measurement of the carrier drift time and, hence,  mobility) could be used for the estimation of the behavior of the correlation function in organic glasses. In practice, experimental complications (noise etc.) make very difficult (and, usually, impossible) to distinguish $E^{1/2}$ and $E^{3/4}$ dependences \cite{Novikov:954}.

\section{Electrostatic energetic disorder near the electrode}
\label{sect_interface}

Structure of the organic material in the vicinity of the electrode is quite different from the bulk structure. We should expect different packing of spacious organic molecules, accumulation of impurities, partial degradation of the organic material etc. For all these reasons the energetic disorder at the interface is very different from the bulk disorder (typically, it is greater than the bulk disorder).

Yet in organic glasses there is an opposite general contribution, leading to the decrease of the disorder at the electrode. Indeed, we already noted that the dominant part of the total energetic disorder in organic glasses is an electrostatic (dipolar or qudrupolar) disorder. The electrostatic energetic disorder is directly proportional to the disorder in the spatial distribution of electrostatic potential, generated by molecular dipoles or quadrupoles. In organic layers sandwiched between  conducting electrodes this spatial distribution must obey a boundary
condition at the electrode surface: here the potential
should be a constant. Thus, at the electrode there
is no electrostatic disorder at all, irrespectively to how disordered
is the material in the bulk. This means that the
magnitude of the dipolar or quadrupolar disorder increases
while going away from the interface, asymptotically reaching its
bulk value.

Some decrease of the disorder at the surface of organic material is inevitable for any model of organic glass (just because there are more neighbor molecules in the bulk of the material), but the magnitude of the effect in the case of electrostatic disorder is much greater than in the case of short range interactions. For example, for the simple model of the interaction with the nearest neighbors only and simple cubic lattice we have $\sigma^2_{\rm surface}=5/6\sigma^2_{\rm bulk}$, while for the dipolar disorder $\sigma^2_{\rm surface}\approx 0.3\sigma^2_{\rm bulk}$ and rms disorder depends on the distance $z$ from the interface as
\begin{equation}\label{sigma2(z)_3}
\sigma^2(z)\approx\sigma^2_{\rm bulk}
\left[1-\frac{a_0}{2z}\left(1-e^{-2z/a_0}\right)\right],
\hskip10pt a_0=Aa,
\end{equation}
and for a simple cubic lattice $A=0.76$ \cite{Novikov:033308}. Spatial correlations at the interface do differ too; this phenomenon has no analogue for the short range disorder. A direct calculation of the correlation function $C(\vec{r})$ near the interface gives
\begin{equation}\label{z=z'}
    C(z_1,z_2,\vec{\rho})=\sigma^2_{\rm bulk} a_0\left(\frac{1}{r_-}-\frac{1}{r_+}\right),
\end{equation}
where $r^2_\pm=\rho^2+(z_1\pm z_2)^2$ and $\vec{\rho}$ is a 2D vector oriented along the interface plane \cite{Novikov:949}. Hence, at the interface the dipolar glass is much less correlated in comparison with the bulk:
$C(z_1,z_2,\vec{\rho})\propto z_1z_2/\rho^3$ for $\rho \gg
z_1,z_2$, and clusters are elongated perpendicular to the interface.

Decrease of the disorder at the interface and change of the spatial behavior of the correlation function have very important implications for the charge injection. In the absence of these phenomena, injection current in organic glasses demonstrates formation of channels where the current density is much greater than the average density \cite{tutis}. Such channels originate from particular spots at the interface, where clusters of sites with low energy
facilitates injection. Reduction of the disorder at the interface and modification of the spatial behavior of the correlation function lead to the more uniform distribution of the injection current over the electrode and dramatically reduce current channeling. This decreases local overheating in a device and improves its performance.

\section{Distribution of cluster size for deep clusters}
\label{sect_clusters}

A common feature of any random medium is the formation of clusters. One of the basic characteristics of the cluster statistics is the cluster distribution on size (or cluster numbers) $n_s$, here $s$ is the number of sites in a cluster.

Analytical results for $n_s$ are scarce, even for the simplest case of non-correlated disorder. The reason for the scarcity is obvious: it is
difficult to take into account various shapes of clusters. Most results
in this area were obtained  using scaling arguments with
subsequent testing of their validity  with extensive computer
simulation \cite{Stauffer:book,Rapaport:679,Grassberger:036101}.

It turns out that a very simple
calculation of the distribution $P_{V}(U_{0})$ of the average value of the
correlated Gaussian field $U_0$ in finite domain having volume $V$
provides  a very accurate estimation of the leading asymptotic
for the distribution of large clusters with $s \gg 1$ and  $U_0
\gg \sigma$.

The distribution $P_{V}(U_{0})$ for the Gaussian random field can be
calculated exactly \cite{Novikov:041139}
\begin{equation}
\label{result}
P_{V}(U_{0}) = \frac{V}{\sqrt{2\pi K}}\exp \left(-\frac{U_0^2 V^2}{2K}\right),
\end{equation}
$$
K = \int_V{d\vec{r}d\vec{r}_1C(\vec{r}-\vec{r}_1)}.
$$
For a spherical domain with radius $R_0$ in the dipolar glass
\begin{equation}
K = \frac{32\pi^2}{15}Aa \sigma^2 R_0^5\propto V^{5/3},
\label{kappa_s}
\end{equation}
while for the non-correlated field $K\propto R_0^3\propto V$ \cite{Novikov:041139}.

We may expect that Eqs. (\ref{result},\ref{kappa_s}) give a reasonable estimation for the number  $n_s$ of the true clusters, i.e. domains, where $U(\vec{r}) > U_0$ everywhere (assuming $V=a^3 s$), at least for the leading term of the asymptotic dependence of $n_s$ on $s$ (the very use of the continuous model of the random medium suggests that our consideration is valid only for $s\gg 1$). If so, then in the dipolar glass
\begin{equation}
\ln n_s \propto -B\frac{U_0^2}{\sigma^2} s^{1/3}, \hskip10pt B=\frac{5}{4A(36\pi)^{1/3}}=0.34...
\label{erfc4}
\end{equation}
This hypothesis agrees well with the simulation data \cite{Novikov:041139}. For the non-correlated Gaussian field it gives a well-known exact asymptotics $\ln n_s\propto -U^2_0 s/2\sigma^2$ \cite{Parisi:871}.

\begin{figure}[floatfix]
\begin{center}
\includegraphics[width=2.8in]{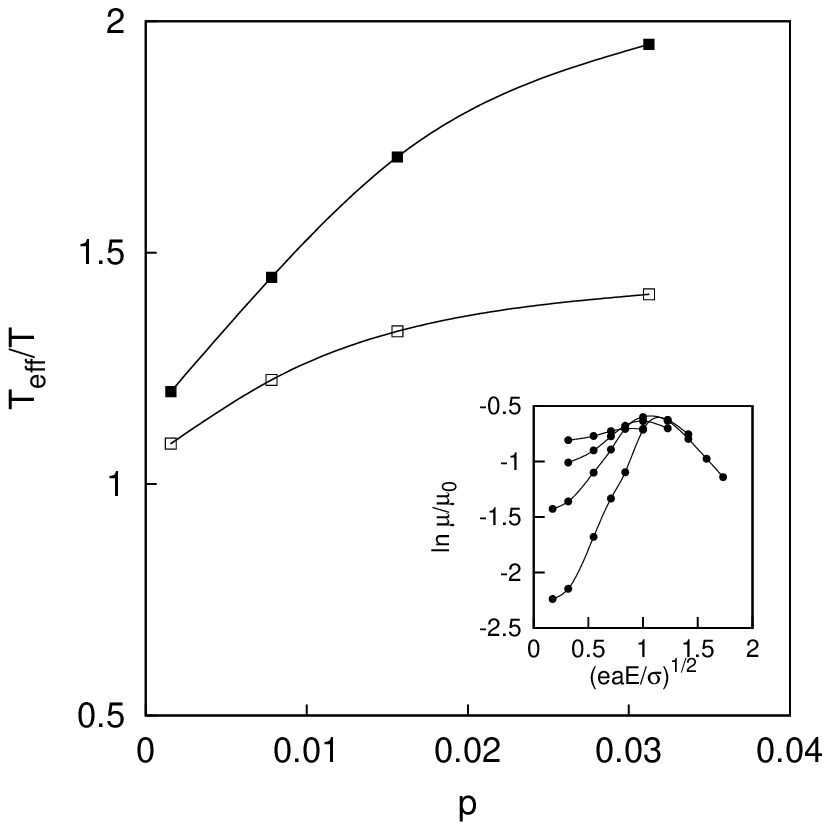}
\includegraphics[width=2.8in]{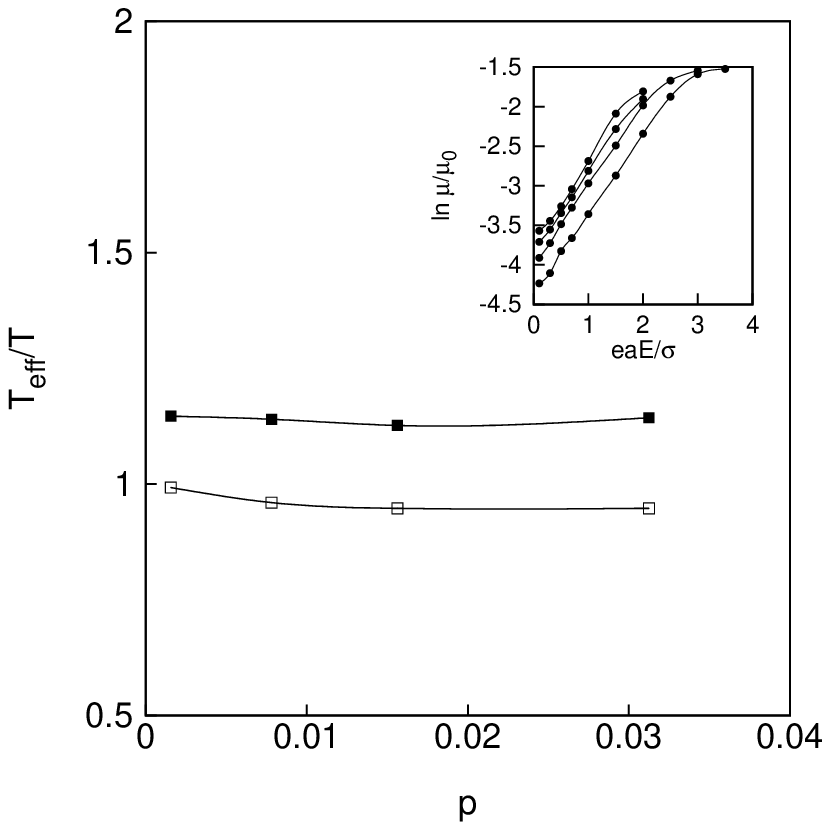}
\end{center}
\caption{Dependence of the effective temperature on the carrier density $p$ for $kT/\sigma=0.3$ ($\blacksquare$) and $kT/\sigma=0.4$ ($\square$) for the DG model (top plot) and GDM (bottom plot), correspondingly. Insets show the curves $\mu(E)$ for $kT/\sigma=0.3$ and different $p$:  0.0016, 0.008, 0.016, and 0.032, from
the bottom curve upward, correspondingly. For the DG $T_{\rm eff}$ was calculated by fitting the mobility data to Eq. (\ref{eq5}), while for the GDM the mobility data was fitted to Eq. (\ref{mu_GDM}). Lines are provided as guides for an eye.}
\label{Teff}
\end{figure}

\section{Charge transport of interacting carriers in Gaussian glasses}
\label{sect_interaction}

If charge carrier density is not very low, we cannot neglect the Coulomb interaction between carriers. The relevant interaction strength parameter is $\lambda=e^2n^{1/3}/\varepsilon\sigma$, where $\varepsilon\simeq 2.5 - 3$ is a dielectric constant of the organic material and $n$ is a carrier concentration. The maximal value of $\lambda$ for $n=1/a^3$ in organic glasses is $\lambda_{\rm max}\simeq 5$. This interaction provides an additional complication to carrier dynamics. Yet spatial correlation manifests itself even in the case of high carrier density.

Indeed, a general tendency for the DG model is that transformation of the mobility
curve with the increase of average fraction of the occupied sites $p=na^3$ resembles the corresponding transformation of the curve with the increase of $T$ (compare insets in Fig. \ref{Teff}, for example, with Fig. 1 in Ref. \cite{Novikov:4472}): with the increase of $p$ mobility becomes greater and the slope of $\mu(E)$ curve becomes smaller. This is not the case for the GDM: here only mobility curve moves upward but the slope remains
approximately constant. Details of the simulation can be found in Ref. \cite{Novikov:740}

This difference could be easily understood. It was noted that the field dependence of
$\mu$ in the GDM is governed by the carrier escape from deep states to the nearest sites having much higher energy, and the field-induced shift of site energies leads to Eq. (\ref{mu_GDM_simple}). Random charge distribution provides a smooth random energy
landscape superimposed on the intrinsic disorder, but typical
additional variation of energy at the scale $a$ is negligible for
small $p$. Hence, estimation (\ref{mu_GDM_simple}) remains valid and
the slope of the mobility curve does not depend on $p$.

Situation in the DG model is different: here mobility field
dependence is governed by the carrier escape from critical clusters, as described in Section \ref{sect_transport}. If we increase the density of
carriers, then at first they fill these critical traps, because
the release time is maximal here. Hence, transport of more mobile
carriers is governed by clusters with the size that
differs from $r_{\rm cr}$ (it is smaller). This means that the effective critical size $r^{\rm eff}_{\rm cr}$ depends on $p$. According to Eq. (\ref{critical}), this is equivalent to the dependence of the effective temperature $T_{\rm eff}$ on $p$: $T_{\rm eff}$ should grow with $p$. This conclusion is in good agreement with Fig. \ref{Teff}: while for the GDM $T_{\rm eff}$ does not depend on $p$ and is very close to $T$, for the DG model $T_{\rm eff}$ monotonously grows with $p$.

\section{Conclusion}
\label{sect_conclusion}

 In organic glasses the dominant part of the random energy landscape is formed by the electrostatic contributions from randomly located and oriented dipoles and quadrupoles. Long range electrostatic contributions automatically provide long range spatial correlations of random energies. These correlations directly dictate the form of the mobility field dependence for the hopping charge transport in amorphous organic materials, even in the case of high carrier density, where carrier-carrier interactions cannot be neglected. Electrostatic disorder is reduced at the interface with the electrode, thus significantly modifying the injection properties. A simple but effective procedure for the calculation of cluster distribution on size for a correlated random Gaussian field is suggested.

\begin{acknowledgement}
This work was supported by the ISTC grant 3718 and RFBR grant
 08-03-00125.
\end{acknowledgement}


\begin{thebibliography}{10}
\bibitem{Bassler:15} H.~B{\"a}ssler, Phys. Status Solidi B \textbf{175}, 15 (1993).
\bibitem{Borsenberger:9} P. M. Borsenberger, E. H. Magin, M. van der Auweraer, and F. C. de Schyver, Phys. Status Solidi A \textbf{140}, 9 (1993).
\bibitem{Novikov:14573} S.V. Novikov and A.V. Vannikov, J.
    Phys. Chem. \textbf{99}, 14573 (1995).
\bibitem{Novikov:361} S.V. Novikov and A.V. Vannikov, Mol.
    Cryst. Liq. Cryst. \textbf{361}, 89 (2001).
\bibitem{Dunlap:437} D.H. Dunlap, V.M. Kenkre, and P.E. Parris, J. Imaging Sci. Tech., \textbf{43}, 437 (1999).
\bibitem{Dunlap:542} D.H. Dunlap, P.E. Parris,
    and V.M. Kenkre, Phys.
    Rev. Lett. \textbf{77}, 542 (1996).
\bibitem{Parris:5295} P.E. Parris, M. K\'{u}s, D.H. Dunlap,
    and V.M. Kenkre, Phys.
    Rev. E \textbf{56}, 5295 (1997).
\bibitem{Novikov:4472} S.V. Novikov, D.H. Dunlap, V.M. Kenkre,
    P.E. Parris, and A.V. Vannikov, Phys. Rev. Lett.
    \textbf{81}, 4472 (1998).
\bibitem{Novikov:954} S.V. Novikov, Annalen der Physik
     \textbf{18}, 954 (2009).
\bibitem{Novikov:033308} S.V. Novikov and G.G. Malliaras, Phys. Rev. B,
    \textbf{73}, 033308 (2006).
\bibitem{Novikov:949} S.V. Novikov, Annalen der Physik
     \textbf{18}, 949 (2009).
\bibitem{tutis} E. Tuti\v{s}, I. Batisti{\'c}, and D. Berner,
Phys. Rev. B \textbf{70}, 161202 (2004).
\bibitem{Stauffer:book}
 D.~Stauffer  and A.~Aharony, Introduction to Percolation Theory
  (Taylor and Francis, London, 1992).
\bibitem{Rapaport:679}
D.C. Rapaport,
  J. Stat. Phys.
  \textbf{66}, 679 (1992).
\bibitem{Grassberger:036101}
P. Grassberger, Phys. Rev. E \textbf{67}, 036101 (2003).
\bibitem{Novikov:041139} S.V. Novikov and M. Van der Awerauer, Phys.
    Rev. E \textbf{79}, 041139 (2009).
\bibitem{Parisi:871}
 G.~Parisi  and
   N.~Sourlas,
   Phys. Rev. Lett. \textbf{46},
   871 (1981).
   \bibitem{Novikov:740} S.V. Novikov, Phys. Status Solidi C
     \textbf{5}, 740 (2008).
\end{thebibliography}
\end{document}